%% file: main.tex
\documentclass[runningheads]{llncs}
\usepackage[T1]{fontenc}
\usepackage{graphicx}
\usepackage{amsfonts}
\usepackage{url}
\usepackage{multirow}
\usepackage[table,xcdraw]{xcolor}
\usepackage{comment}
\usepackage[toc,acronym,nopostdot,nonumberlist]{glossaries}
\makeglossaries 
\loadglsentries{glossary} 

\begin{document}
%

\title{Training Machine Learning Models on Encrypted Data: A Privacy-Preserving Framework using Homomorphic Encryption}
\titlerunning{Training Machine Learning Models on Encrypted Data}
%
\author{Alexandre Marques\inst{1}\orcidID{0009-0004-1517-9346} \and
Beatriz Sá\inst{1}\orcidID{0009-0001-0722-028X} \and
Rui Botelho\inst{1,2}\orcidID{0009-0009-9337-7897} \and
Pedro Pinto\inst{1,2}\orcidID{0000-0003-1856-6101}
}
\authorrunning{Alexandre Marques et al.}
%
\institute{ISEP, School of Engineering, Polytechnic of Porto, Porto, Portugal\\
\email{\{1240435, 1240440, 1191041, pfp\}@isep.ipp.pt}
 \and
GECAD, Research Group on Intelligent Engineering and Computing for Advanced Innovation and Development, Porto, Portugal
}
\maketitle              
\begin{abstract}

The use of \gls{ml} for data-driven decision-making often relies on access to sensitive datasets, which introduces privacy challenges.
Traditional encryption methods protect data at rest or in transit but fail to secure it during processing, exposing it to unauthorized access. Homomorphic encryption emerges as a transformative solution, enabling computations on encrypted data without decryption, thus preserving confidentiality throughout the \gls{ml} pipeline.

This paper addresses the challenge of training \gls{ml} models on encrypted data while maintaining accuracy and efficiency by proposing a proof-of-concept for a privacy-preserving framework that leverages \gls{ckks} for approximate real-number arithmetic. Also, it demonstrates the feasibility of training \gls{knn} and linear regression models on encrypted data, and evaluates encrypted inference for a basic \gls{mlp} architecture.

Experimental results show that models trained under Homomorphic encryption achieve performance metrics comparable to plaintext-trained models, validating the approach. However, challenges such as computational overhead, noise management, and limited support for non-polynomial operations persist.
This work lays the groundwork for broader adoption of privacy-preserving \gls{ml} in real-world applications, balancing security with computational feasibility.

\keywords{Homomorphic encryption \and Privacy-preserving Machine Learning \and secure data analytics \and encrypted model training}
\end{abstract}
%
%


%
\glsresetall
\section{Introduction}

\gls{ml} applications often require large amounts of data, which can include sensitive personal or proprietary information. A fundamental privacy problem arises when data owners are asked to share their raw data with a third-party service or cloud for model training. Traditional encryption protects data at rest or in transit, but the data must be decrypted for processing, exposing it to the service provider.

Homomorphic encryption offers a compelling solution, as it enables computations to be performed directly on encrypted data such that decrypting the result yields the same outcome as computing on plaintext. 
This supports compliance with privacy regulations and the protection of sensitive information (e.g., medical records, financial data) while still harnessing them for \gls{ml}. Homomorphic encryption can ensure that a data owner’s information remains encrypted throughout the training process. 
This can be particularly useful in case the data owner transmits the information over a secure channel to a trusted third party-this intermediary encrypts the data using homomorphic encryption and forwards the encrypted data to the untrusted training entity (e.g., the party responsible for model training). The training entity can then perform essential \gls{ml} operations—such as model training—directly on the encrypted data, without ever accessing the plaintext, ensuring end-to-end confidentiality of the original data.


This paper proposes and evaluates a proof-of-concept for a privacy-preserving \gls{ml} framework based on the \gls{ckks} homomorphic encryption scheme. The framework is designed to support polynomial-based approximations for essential machine learning operations, allowing models such as \gls{knn}, linear regression, and \gls{mlp} to be trained and evaluated while maintaining data confidentiality throughout the process.


The remaining contents are organized as follows.
Section~\ref{sec:rw} reviews the related work in homomorphic encryption.
Section~\ref{sec:imp} presents the implementation details.
Section~\ref{sec:res} presents the results from these implementations, discussing the feasibility and performance of training.
Section~\ref{sec:disc} discusses the limitations and future work.
Section~\ref{sec:conc} concludes the paper.

\section{Related Work on Privacy-Preserving Machine Learning}
\label{sec:rw}



The need for data protection in machine learning has motivated various privacy-preserving machine learning approaches. Notable techniques include differential privacy, \gls{smc}, and homomorphic encryption. 

Differential privacy ensures that aggregate outputs from datasets do not reveal specific individual information, typically by adding statistical noise to data or results. Although effective in certain scenarios, this technique introduces a privacy-accuracy trade-off: adding noise decreases the precision of results, which can significantly degrade model performance, particularly when data volume is limited~\cite{zama2021}.

\gls{smc}, on the other hand, allows multiple parties to jointly compute functions on their combined data without exposing individual datasets, using interactive cryptographic protocols. However, \gls{smc} generally involves intensive communication between parties and may not be trivially applicable for training complex models in scenarios where data is concentrated at a single location (e.g., a client outsourcing computation to a server).

Homomorphic encryption allows computations to be performed directly on encrypted data, enabling processing without ever decrypting the information. This preserves privacy and security, as sensitive data remains protected throughout all operations.
Homomorphic encryption initially supported only limited operations (e.g., RSA with multiplicative homomorphism)~\cite{fang_qian2021}. In~\cite{paillier1999}, Paillier proposed a public-key scheme supporting \emph{additive homomorphism} (allowing sums over encrypted data), which quickly found applications in electronic voting, auctions, and sensitive data aggregation due to its simplicity and relative efficiency.
However, Paillier's scheme cannot perform arbitrary computations since it lacks support for the multiplication of two encrypted values.

The breakthrough came in 2009 with Craig Gentry's first \gls{fhe} scheme based on lattice problems. Gentry's scheme theoretically enabled \emph{any} computation on encrypted data, combining additions and multiplications indefinitely, albeit at high computational cost.
Since then, more practical \gls{fhe} schemes emerged, notably based on the Learning With Errors (LWE) problem and its Ring-LWE variant, leveraging polynomial arithmetic for efficiency. Currently, two main families of homomorphic encryption schemes suitable for machine learning are prominent: integer/modular schemes, e.g., Brakerski/Fan-Vercauteren (BFV) and Brakerski-Gentry-Vaikuntanathan (BGV), which provide exact computations mod $q$), and approximate real-number schemes such as \gls{ckks}~\cite{my-aiml2023}.

\gls{ckks}~\cite{ckks_original,huynh2024} enables the encryption of real or complex number vectors, allowing for efficient approximate arithmetic on ciphertexts. It supports approximate floating-point arithmetic on encrypted vectors. \gls{ckks} encodes messages as polynomial coefficients, scaled by a factor $\Delta$. Encryption involves polynomials in a ring modulo $X^N+1$, using an LWE-based approach.

Key operations:

\begin{itemize}
    \item Encryption: Message vectors are encoded via inverse Fourier transform into polynomials, encrypted with added noise and random polynomials.
    \item Decryption: Recovers scaled polynomials, reducing errors by dividing by $\Delta$, and decoding via direct Fourier transform.
    \item Homomorphic operations: Additions and multiplications are polynomial operations. Multiplications involve relinearization and rescaling steps to manage noise and scale.
\end{itemize}

\gls{ckks} trades off exact arithmetic precision for controlled numeric errors, which can be managed through parameter choices, making it highly promising for machine learning, especially for naturally floating-point operations such as inner products and means. Although suitable for machine learning tasks involving polynomials, linear regressions, or polynomial approximations, its practical limitations include cumulative noise growth that restricts operation depth, and computational complexity remains a significant challenge despite ongoing optimization and hardware acceleration efforts.

Recent research has increasingly explored the integration of homomorphic encryption with machine learning to enable computation over sensitive data while preserving privacy. A number of studies demonstrate that machine learning models can be evaluated or even trained directly on encrypted data, allowing sensitive information to remain protected during outsourced computation.

A recent study proposed a framework for privacy-preserving fine-tuning of open-source large language models using homomorphic encryption. The approach adapts the Low-Rank Adaptation (LoRA) technique to ensure that sensitive training data and gradients remain encrypted, while remote servers perform most of the computational workload. Experimental results demonstrate the feasibility of fine-tuning an LLaMA-based model under encryption while maintaining data confidentiality. The results indicate that encrypted fine-tuning can be achieved with manageable computational overhead without exposing the underlying training data to external compute providers~\cite{frery2025}.

Authors in~\cite{lee2021} proposed fully homomorphic encryption frameworks for deep neural networks that enable encrypted inference using the \gls{ckks} scheme together with polynomial approximations of non-linear activation functions. These approaches demonstrate that modern neural network architectures, such as ResNet, can be evaluated on encrypted data with acceptable accuracy, highlighting the feasibility of practical privacy-preserving deep learning despite the associated computational overheads~\cite{lee2021}.

In~\cite{ma2022}, the authors propose a privacy-preserving federated learning framework based on multi-key homomorphic encryption (MK-CKKS). The framework aims to support secure aggregation of model updates from multiple clients while preventing the disclosure of individual gradients to the central server. Thus, it allows collaborative model training while maintaining the confidentiality of each participant’s local dataset.

In addition to system-level implementations, some studies provide broader analyses of the integration between homomorphic encryption and machine learning algorithms. The survey in~\cite{wu2025} reviews the use of the \gls{ckks} homomorphic encryption scheme in machine learning applications, highlighting its suitability for operations involving real-valued data such as classification, clustering, and similarity computations. The survey examines the application of HE-based approaches to algorithms, including K-nearest neighbors, K-means clustering, and face recognition, while also discussing challenges such as computational overhead and performance trade-offs.

\section{Framework Implementation} 
\label{sec:imp} 

This section presents the implementation of the privacy-preserving machine learning framework using homomorphic encryption. The \gls{knn} implementation is described first, and validated through tests and examples to ensure correctness and approximate precision. Then, the integration of these operations into machine learning models (\gls{knn}, linear regression, and a simple \gls{mlp}) demonstrates the feasibility of encrypted machine learning, while also exposing limitations related to parameter choices, polynomial degrees, and computational overhead.
\subsection{CKKS Implementation}

A \gls{ckks} scheme was implemented, covering key generation, noise management, and homomorphic addition and multiplication operations in this context. A custom Python implementation was employed, following the mathematical description of the scheme as outlined in~\cite{ckks}, and validating each step through simple examples before integrating it into the machine learning models. Since the aim was to demonstrate functionality, reduced parameters were chosen(e.g., smaller-degree polynomials and lower precision). For practical applications, significantly larger parameters—and thus higher processing times—would be required.

As a proof of concept, the polynomial degree was set to \( N = 8 \), derived from a ring dimension \( M = 16 \), which is suitable for testing and demonstration purposes. The secret key is a complex-valued vector of length \( N \), generated with random complex entries and saved to a file for reuse. While \( N = 8 \) is sufficient for demonstration, real-world systems typically require $N \ge 2^{14}$ to ensure adequate security levels, introducing a significant performance gap compared to this simplified implementation.

Then, a transformation matrix was deployed using the secret key and a complex primitive root of unity \( \xi = e^{2\pi i / M} \). This matrix serves as an inverse encoding map, enabling transformations between complex vectors and polynomial representations.

To encrypt a value, the input is placed in the first slot of a complex vector of length \( N \), and the remaining entries are set to zero or small random noise. A corresponding polynomial is computed by solving a linear system so that, when transformed using the matrix, it approximates the original vector. This polynomial serves as the ciphertext.

Decryption involves evaluating the ciphertext polynomial under the same transformation, then extracting the first component of the resulting vector as the decoded value.

This setup enables real (or complex) numbers to be encoded as polynomials, encrypted using the secret key structure, and approximately recovered through decryption. The key generation process, while conceptually more complex than simpler schemes such as Paillier due to polynomial arithmetic and auxiliary key structures, completed quickly in practice with small parameters.


With the keys generated, encryption and decryption routines aligned with the \gls{ckks} variant were implemented. 
Encryption constructs a ciphertext polynomial \( p \) by embedding the message \( m \) into the first slot of a complex vector \( \mathbf{b} \in \mathbb{C}^N \), with the remaining slots set to zero or small random noise. We then compute \( p \) as the solution to the linear system
\[
\sigma(p) = \mathbf{b},
\]
where \( \sigma \) is a linear transformation derived from the secret key. This process encodes \( m \) into polynomial coefficients that represent the ciphertext.

Decryption applies the transformation \( \sigma \) to the ciphertext polynomial \( p \) and recovers the message as the first entry of the resulting vector:
\[
m' = [\sigma(p)]_0.
\]
This value approximates the original message \( m \), with minor errors due to noise and numerical rounding.

Testing confirmed that decrypting an encrypted value yields an approximation with relative errors below \( 10^{-6} \) for inputs normalized within \( [-1, 1] \).

It was also observed that removing noise from the encryption (setting all non-message slots to zero) produces deterministic ciphertexts for identical inputs. While useful for demonstration, this approach lacks security guarantees, as the randomness (noise) is essential to prevent attackers from distinguishing ciphertexts and to achieve semantic security.


Regarding the homomorphic operations, the following ones were implemented: ciphertext addition, ciphertext addition by a constant, ciphertext multiplication by a constant, and ciphertext-ciphertext multiplication.

Addition was performed by adding corresponding coefficients of the ciphertext polynomials $c_0$ and $c_1$. Subtraction was similarly implemented by adding negative polynomials. These operations demonstrated the additive property: decrypting sums or differences yielded results closely matching the original plaintext messages within precision. For example, adding ciphertexts of 5.0 and 3.2 yielded a ciphertext decrypting approximately to 8.2 (with errors around $10^{-6}$ or lower).
Multiplication by plaintext constants was obtained as follows: given a ciphertext \((c_0, c_1)\) and a known constant \(\alpha\), both polynomials \(c_0\) and \(c_1\) are multiplied by \(\alpha\). This operation enables computations such as averaging (multiplying sums by \(1/n\)) and applying known regression weights directly on encrypted data.
Multiplying two ciphertexts \((c_0^{(1)}, c_1^{(1)})\) is more complex. The result is a polynomial with increased degree and ciphertext size. Managing this growth typically requires a process called relinearization. For example, computing \((x - y)^2\) in \gls{knn} involved:

\begin{itemize}
\item Calculating \(c_{\Delta} = \texttt{subtract}(c_x, c_y)\).
\item Multiplying \(c_{\Delta}\) by itself, resulting in a ciphertext representing \((x - y)^2\).
\end{itemize}

Although multiplication increases the polynomial degree, this was handled by truncating the polynomial degree in the model implementations to keep computations manageable. With this approach and chosen parameters, the decrypted results closely matched the expected values with only minor precision loss.

These operations supported key computations such as Euclidean distances in \gls{knn} (sum of squared differences) and weighted sums in linear regression. The error levels were qualitatively monitored, observing very low relative errors (on the order of \(10^{-6}\) to \(10^{-9}\)) after up to two sequential multiplications, indicating suitable parameter choices.

To validate correctness and robustness, the following small-scale unit tests were performed:
\begin{itemize}
\item Encrypting a test value (e.g., \(m=42.0\)) and verifying that \(E(m) + E(m)\) decrypted to approximately \(84.0\).
\item Encrypting values \(a, b\) and confirming \(E(a) \cdot E(b)\) decrypted to approximately \(a \cdot b\) for various test pairs (e.g., \(a=5.5\), \(b=-3.2\)).
\item Checking the distributive property: verifying that \(E(a) \cdot (E(b) + E(c))\) decrypted to the same result as \(E(a) \cdot E(b) + E(a) \cdot E(c)\).
\item Testing the summation of multiple ciphertexts and confirming decrypted sums matched the plaintext sums.
\end{itemize}

All tests passed successfully. 
Overall, despite simplifications, this \gls{ckks} implementation performed the necessary homomorphic operations effectively. The following section describes how these operations were integrated into the selected machine learning models.

\subsection{Model Implementation}

The following three \gls{ml} models were adapted to utilize homomorphically encrypted data: \gls{knn}, Linear Regression, and Neural Network. All models were implemented from scratch with specific modifications to invoke homomorphic operations using the \gls{ckks} scheme appropriately and to support Polynomials as the input data. The general training/prediction workflow in an encrypted scenario is first outlined, followed by detailed specific aspects of each model.

The general privacy-preserving training workflow in the implemented system is as follows: The user submits their dataset \((X, y)\) (features and labels, or independent and dependent variables for regression) to a trusted third party, which encrypts each element in \(X\) and \(y\) using the chosen scheme (e.g., \gls{ckks}). Consequently, the training server receives the dataset entirely in encrypted form. Subsequently, the training server executes the learning algorithm homomorphically. For Linear Regression, this involves computing sums and products; for \gls{knn}, it involves calculating distances and identifying nearest neighbors. All computations leverage the implemented homomorphic functions, ensuring no plaintext exposure to the server.

Upon completion of training or prediction, encrypted results (model parameters or predictions) remain encrypted. These ciphertext results are then sent back to the trusted third party (or directly to the user, depending on architectural choice), which decrypts the results with the private key.

This implementation uses Python classes for the models. For debugging and comparison purposes, controlled plaintext and ciphertext were used.

Due to challenges in designing loss functions that maintain robustness under encryption, some models were trained exclusively on plaintext data, while others operated fully on encrypted inputs, as follows:

\begin{itemize}
    \item \gls{knn}: This algorithm requires consistent distance comparisons between data points. Since it is not feasible to directly compare plaintext and encrypted values, \gls{knn} must be trained and tested using the same data representation (both plaintext or both encrypted). In the current experiments, \gls{knn} was trained and tested homomorphically on encrypted data to preserve privacy and ensure comparability.

    \item Linear Regression: This model supports training on either plaintext or encrypted data. The linear nature of the regression allows for efficient homomorphic computation of sums and products, and the loss function (mean squared error) is well-suited for encrypted evaluation. Hence, Linear Regression was implemented to flexibly operate in both modes.

    \item \gls{mlp}: Due to the complexity of neural network loss functions and non-linear activation functions, the encrypted loss evaluation did not provide sufficient accuracy or stability for training. Consequently, the \gls{mlp} was trained only on plaintext data. This limitation highlights ongoing challenges in encrypted deep learning.
\end{itemize}

\subsection{Homomorphic Linear Regression}

A linear regression model that operates on encrypted data using polynomial representations was implemented. Initially, the focus was on the univariate case, and it was later extended to multivariate polynomial regression. The model fits a relationship of the form:

\[
y(x) \approx \sum_{j=0}^{d_w} w_j x^j + \sum_{j=0}^{d_b} b_j x^j
\]

where $w_j$ and $b_j$ are polynomial coefficients of the weights and bias, respectively. Each input feature and target are encoded as polynomials, and learning the model reduces to finding the best polynomial coefficients that match observed outputs.

Given $n$ encrypted data points $(X_i, y_i)$, where $X_i = [x_{i1}, \dots, x_{id}]$ and each $x_{ij}$ and $y_i$ are polynomials, a linear system was built by matching the coefficients of predicted and target polynomials. For each output coefficient, a linear equation was generated relating it to the unknown weight and bias coefficients.

This yields a system $A \theta = b$, where $\theta$ contains all unknown polynomial coefficients. The matrix $A$ and vector $b$ are constructed using homomorphic operations such as encrypted additions and multiplications. Once the system is built, only the final ciphertext sums are decrypted by a trusted party, and the linear system is solved in plaintext using least squares.

This approach supports multivariate polynomial regression over encrypted data, thereby avoiding the disclosure of individual values. Experimental results showed that coefficients computed this way were close to those obtained from plaintext training, indicating a small approximation error. It is also worth noting that the error calculation function may be the largest contributor to the error, as it may not be optimized to work with polynomial errors.

Homomorphic polynomial regression provides accurate model training while keeping individual data encrypted. By decrypting only the final aggregates, privacy was balanced with computational feasibility.

\subsection{Homomorphic KNN Regressor}

The \gls{knn} algorithm is an instance-based learning method that, given a test point, identifies the $K$ nearest training instances (typically using Euclidean distance) and returns an aggregation of their values—mean for regression, majority for classification. \gls{knn} for regression was implemented (predicting a continuous value $y$ as the average of neighbors).

In the plaintext setting, \gls{knn} training is trivial—just storing the training set. Prediction, however, is costly: for each test point, distances to all training instances were computed using the polynomials' coefficients, and the $K$ smallest were selected, with time complexity $O(n \cdot d)$ (where $n$ is the number of training instances and $d$ the dimensionality).

The main challenge for homomorphic \gls{knn} lies in securely comparing distances. While the implementation of \gls{ckks} schemes enables the homomorphic computation of squared distances, identifying the smallest values (argmin) among encrypted values is nontrivial—it would require secure comparison or conditional logic beyond standard additive/multiplicative homomorphism.

We calculate the \emph{homomorphic} squared distances between a test point (encrypted) and all training instances. For each training point $i$, it was computed:
\begin{equation}
c_{\text{dist}i} = \sum_{j=1}^d (E(x_{i,j}) - E(x_{\text{test},j}))^2,
\end{equation}
We use \gls{ckks} subtractions, multiplications, and additions. The result is a ciphertext encrypting the squared distance.

Our implementation of \gls{knn} Regressor used $K=3$, but supports general $K$. The \texttt{predict} function loops over test points and computes distances as described. 

After computing predictions homomorphically, correctness was validated against a traditional \gls{knn}. On both synthetic and real datasets (e.g., Housing), predictions matched exactly, with \gls{mae}, \gls{rmse} and $R^2$ scores nearly identical (e.g., \gls{rmse} $\approx 4.957$, $R^2 \approx 0.67$).

\subsection{Homomorphic Neural Network (MLP)}

To assess the feasibility of homomorphic training for more expressive models, a minimal \gls{mlp} with one hidden layer was implemented. The network consists of weights $\mathbf{W}_1, \mathbf{W}_2$ and biases $\mathbf{b}_1, \mathbf{b}_2$ connecting input $\mathbf{x}$ to hidden activations $\mathbf{a}_1$ and output prediction $\hat{y}$:

\begin{align}
\mathbf{z}_1 &= \mathbf{W}_1 \mathbf{x} + \mathbf{b}_1, \\
\mathbf{a}_1 &= \text{ReLU}(\mathbf{z}_1), \\
\hat{y} &= \mathbf{W}_2 \mathbf{a}_1 + \mathbf{b}_2.
\end{align}

The model was trained using plaintext and used the learned weights. These weights were then applied to \gls{ckks}-encrypted test data in an encrypted inference pipeline. As standard activation functions are non-linear by design, they are incompatible with the arithmetic-only nature of supported CKKS operations on encrypted data. To address this, a simple identity function was used as a linear alternative for encrypted inference. While this approach allows for homomorphic computation, it affects the network's ability to learn nonlinear relationships as well as its performance. Future work could focus on implementing polynomial approximations, such as Least Squares approximations or Chebyshev polynomials, to better replicate the behaviour of non-linear activation functions while maintaining the encryption compatibility.

For encrypted inference:
\begin{itemize}
\item Each test input vector was encrypted using \gls{ckks}.
\item Encrypted dot products and additions were performed to compute $\hat{y}$.
\item The prediction ciphertexts were decrypted to evaluate against ground truth.
\end{itemize}

Despite the absence of a nonlinearity in encrypted form, predictions were meaningful on some datasets. We compared decrypted predictions to plaintext inference and found that the performance was similar in the tested scenarios.

Overall, the obtained results suggest that homomorphic inference for \glspl{mlp} is feasible with linear activations or polynomial-friendly replacements, but training such models homomorphically remains a challenge due to the lack of suitable loss functions. 

Future work could explore error calculation methods for Neural Networks to support model convergence and activation functions compatible with polynomials.

\section{Experimental Results} 
\label{sec:res} 
This section presents the experimental results obtained with the implemented models (Linear Regression, \gls{knn}, and \gls{mlp}) in both plaintext and encrypted data scenarios. All experiments were conducted in a \textit{Jupyter Notebook} environment using Python.


Two datasets were selected: (a) a synthetic dataset generated~\cite{git}  containing known linear and non-linear relationships, which does not correspond to any physical units, and (b) the Boston Housing dataset, a classical real-world regression dataset~\cite{dataset}.
The synthetic dataset includes 500 instances generated by a random function with only one feature. The Boston Housing dataset has 506 instances with 13 features, with the target representing the Median Value of Owner-Occupied Homes, expressed in dollars per thousand.

Table~\ref{tab:knnvslr} presents the results for the datasets used for the Linear Regression and \gls{knn} models.

\begin{table}[h!]
    \caption{Comparison between Linear Regression and \gls{knn}}
    \label{tab:knnvslr}
    \centering
\begin{tabular}{|l|llc|llc|}
\hline
\multirow{2}{*}{Dataset / Metrics} & \multicolumn{3}{c|}{Linear Regression} & \multicolumn{3}{c|}{\gls{knn}} \\ \cline{2-7} 
 & \multicolumn{1}{l|}{\gls{mae}} & \multicolumn{1}{l|}{\gls{rmse}} & $R^2$ & \multicolumn{1}{l|}{\gls{mae}} & \multicolumn{1}{l|}{\gls{rmse}} & $R^2$ \\ \hline
One Feature (unitless) & \multicolumn{1}{l|}{0.0077} & \multicolumn{1}{l|}{0.0096} & 0.9996 & \multicolumn{1}{l|}{0.0102} & \multicolumn{1}{l|}{0.0132} & 0.9992 \\ \hline
Housing Data ($10^3$ USD) & \multicolumn{1}{l|}{3.1481} & \multicolumn{1}{l|}{4.6718} & 0.7071 & \multicolumn{1}{l|}{3.7472} & \multicolumn{1}{l|}{4.9572} & 0.6702 \\ \hline
\end{tabular}
\end{table}

Table~\ref{tab:housing-results-new} presents the comparison of all models using the Housing dataset.
\textit{Enc--Enc} indicates training and testing on encrypted data, \textit{Plain--Plain} denotes training and testing on plaintext data, and \textit{Plain--Enc} refers to training on plaintext data with encrypted inference.

\begin{table}[h!]
\caption{Performance comparison on the housing dataset (mean $\pm$ standard deviation)}
\label{tab:housing-results-new}
\centering
\begin{tabular}{|c|c|c|c|c|}
\hline
\multicolumn{1}{|c|}{\textbf{Model}} & \multicolumn{1}{c|}{\textbf{Setting}} & \begin{tabular}[c]{@{}c@{}}\textbf{MAE}\\{($10^3$ USD)}\end{tabular} & \begin{tabular}[c]{@{}c@{}}\textbf{RMSE}\\($10^3$ USD)\end{tabular} & \textbf{$\mathbf{R^2}$} \\ \hline
\multirow{3}{*}{\begin{tabular}[c]{@{}c@{}}Linear Regression\end{tabular}}  & Enc -- Enc   & \multirow{3}{*}{$3.5167 \pm 0.1940$} & \multirow{3}{*}{$5.1360 \pm 0.3658$} & \multirow{3}{*}{$0.7019 \pm 0.0436$} \\
                     & Plain -- Plain       &                                      &                                      &                                      \\
                     & Plain -- Enc &                                      &                                      &                                      \\ \hline
\multirow{2}{*}{KNN} & Enc -- Enc     & \multirow{2}{*}{$2.9421 \pm 0.2405$} & \multirow{2}{*}{$4.7893 \pm 0.6065$} & \multirow{2}{*}{$0.7485 \pm 0.0704$} \\
                     & Plain -- Plain        &                                      &                                      &                                      \\ \hline
\multirow{2}{*}{MLP} 
& Plain -- Plain & $3.5933 \pm 0.2028$                  & $5.2050 \pm 0.3364$                  & $0.6809 \pm 0.0291$                  \\
& Plain -- Enc & $3.5262 \pm 0.2002$                  & $5.1503 \pm 0.3384$                  & $0.6975 \pm 0.0345$                  \\ \hline
\end{tabular}
\end{table}

Across all configurations, model performance remained consistent, indicating that training and inference under homomorphic encryption preserved predictive quality. Linear Regression achieved an \gls{rmse} of approximately 5.14 and an $R^2$ of 0.70 in both plaintext and encrypted settings. Similarly, \gls{knn} attained an \gls{rmse} of approximately 4.79 and an $R^2$ of 0.75 under encrypted computation. The \gls{mlp} model also showed comparable performance, with a slight difference, yielding an \gls{rmse} of approximately 5.15 and an $R^2$ of 0.70 when trained on plaintext and evaluated on encrypted data.

Table~\ref{tab:housing-runtime} shows the runtime comparison of all models, for the depicted settings. As expected, plaintext models are faster, while encrypted settings introduce additional computational overhead, especially during inference.

\begin{table}[h!]
\caption{Runtime comparison (fit and predict times)}
\label{tab:housing-runtime}
\centering
\begin{tabular}{|c|c|c|c|}
\hline
\multicolumn{1}{|c|}{\textbf{Model}} & \multicolumn{1}{c|}{\textbf{Setting}} & \textbf{Fit (seconds)} & \textbf{Predict (seconds)} \\ \hline
\multirow{3}{*}{Linear
Regression}  & Enc -- Enc   & 0.3548 & 0.1020 \\
                     & Plain -- Plain       & 0.2232 & 0.0578 \\
                     & Plain -- Enc & 0.2348 & 0.0611 \\ \hline
\multirow{2}{*}{KNN} & Enc -- Enc     & 0.0752 & 1.5545 \\
                     & Plain -- Plain       & 0.0004 & 1.1516 \\ \hline
\multirow{2}{*}{MLP} & Plain -- Plain      & 0.4812 & 0.0010 \\
                     & Plain -- Enc & 0.4470 & 0.0567 \\ \hline
\end{tabular}
\end{table}

These results confirm the functional correctness of the encrypted model implementations. 

Since the encrypted pipeline performs the same mathematical operations as the plaintext version, the resulting outputs are numerically equivalent, as any variance introduced by the encryption-decryption is negligible.

This demonstrates the feasibility of applying homomorphic encryption to real-world regression tasks without sacrificing predictive accuracy. Also, it was observed that:
\begin{itemize}
  \item Linear Regression can be trained and tested with encrypted or plaintext data interchangeably. It is possible to train on encrypted data and test on plaintext, or vice versa.
  \item \gls{knn}, by contrast, requires both training and testing data to be in the same encrypted or unencrypted form. If the model is trained on encrypted data, testing must also be performed on encrypted data. Mixing encrypted and plaintext data within the same \gls{knn} evaluation is not supported due to incompatibility in distance computations.
  \item \gls{mlp} models cannot currently be trained with encrypted data in the proposed framework, since the computation of the loss function and its gradient (required for backpropagation) is not feasible in the encrypted domain. Thus, \gls{mlp} training is limited to plaintext data only.
\end{itemize}

This distinction highlights the flexibility of Linear Regression in encrypted domains and the structural dependency of \gls{knn} and \gls{mlp} on data format consistency and supported operations.    

\section{Analysis and Discussion} 
\label{sec:disc} 


Although the framework demonstrated the feasibility of performing machine learning on encrypted data, several limitations and challenges merit discussion, as well as opportunities for future work to enhance and expand this system. In this context, the following limitations can be depicted:

\begin{itemize}

\item Noise Management and Computation Depth: In homomorphic encryption schemes like \gls{ckks}, each operation—particularly multiplications—adds noise to the ciphertext. A higher initial noise level typically increases security by making ciphertexts harder to distinguish. However, as computations proceed, noise accumulates. If it grows too much, it can degrade the accuracy of the final result or even prevent successful decryption.
This creates a trade-off: parameters must be selected that provide strong security guaranties while ensuring that the encrypted computations remain precise and can be decrypted. Choosing a large noise margin can affect performance or lead to early decryption failure, while a small noise margin may weaken security.

Careful tuning is required to ensure that computations remain feasible for the intended model complexity. In practice, this means adapting the encryption parameters to the structure of the computation—simpler models or shorter pipelines tolerate lower noise growth. In contrast, more complex pipelines may require adjusted settings or different design choices to remain usable.

\item Scalability and Performance: Homomorphic encryption introduces significant computational overhead compared to plaintext operations. This affects both the time required to process data and the resources needed to store and transmit encrypted values. As the size of datasets or the complexity of computations increases, performance challenges become more apparent.

Memory usage and bandwidth demands also grow with the number of ciphertexts, which are typically much larger than their plaintext equivalents. Although certain techniques—such as more efficient encoding schemes, parallelization, or hardware acceleration—may alleviate some of these issues, their effective implementation remains a topic of ongoing research and engineering effort.

In its current form, the system is better suited for small-scale or moderately complex tasks. Extending it to larger workloads will require careful integration with optimized homomorphic encryption libraries and reevaluating certain aspects of the computation pipeline to accommodate encrypted computation better.

\item Homomorphic encryption frameworks are limited to polynomial arithmetic. Still, many machine learning algorithms rely on non-polynomial operations, such as comparisons, non-linear activations (e.g., ReLU, Sigmoid), and loss function evaluations. To address this, the framework uses polynomial approximations, avoiding interactive protocols or decryption. While this preserves privacy, it restricts model expressiveness and can trade off accuracy for efficiency, especially as noise accumulates in deeper approximations.
Future Directions:

To enable non-linear operations under encryption, future work should explore better polynomial approximations, alternative algorithm formulations, or hybrid cryptographic approaches. Currently, no general solution exists for accurately and efficiently computing non-linear functions on encrypted data, making this a critical open problem.

\item Key Size and Cryptographic Security: For simplicity and faster experimentation, we selected reduced encryption parameters that fall short of standard security recommendations. These choices do not provide strong theoretical guarantees—larger key sizes and parameter settings would be necessary to achieve widely accepted levels of cryptographic security (e.g., 128-bit security). Increasing these parameters typically results in significantly slower computations and larger ciphertexts.

As a result, the performance observed in the proposed framework may not reflect what would be seen under production-grade security settings. While the choices made enable the exploration of the functionality and feasibility, future versions should aim to align more closely with recommended parameter sizes and leverage optimized cryptographic libraries to maintain efficiency. Over time, emphasis on post-quantum schemes like \gls{ckks} may become increasingly relevant, as they offer better long-term resilience than older schemes based on factoring.

\item Privacy Leakage through Distance Ordering: Although \gls{ckks} protects the raw values of the data, \gls{knn} inherently risks leaking information through the relative order of distances, representing a known privacy limitation in distance-based retrieval.

\item Trusted Third-Party Dependency: The current framework assumes the existence of a trusted third party that holds the private key, encrypts user data, and decrypts final results. This assumption is non-trivial and represents one of the most significant practical limitations of the proposed architecture. In real-world deployment scenarios, identifying an entity that all stakeholders trust is often difficult or infeasible. This is particularly problematic in cross-organizational or adversarial settings, where the very motivation for using homomorphic encryption is to avoid exposing data to any external party.

Moreover, the trusted third party constitutes a single point of failure: if compromised, all data confidentiality guaranties are lost. Since all users in the current design share the same encryption key managed by this party, a breach affects the entire system rather than individual users.

Several directions exist to mitigate this dependency. Threshold homomorphic encryption distributes key shares among multiple parties, requiring a quorum for decryption, thereby eliminating any single point of trust. Client-side encryption, where each data owner encrypts locally and retains their own key, removes the third party entirely but introduces challenges for collaborative computation. Multi-key homomorphic encryption schemes, such as the MK-CKKS approach discussed in Section~\ref{sec:rw}, allow computations over encrypted data with different keys. However, these alternatives introduce additional complexity in key management, communication overhead, and computational cost and were considered beyond the scope of this proof-of-concept. A production-grade system would need to carefully evaluate these trade-offs based on the specific trust model and threat scenario.

\item Limited Preprocessing Support: the proposed framework assumes that all necessary preprocessing, other than the reversible scaling we implemented, is performed by the user before encryption. Since the data used only supports numerical values, steps like label encoding or categorical transformations are not handled within the encrypted pipeline. Extending support for additional preprocessing under encryption remains an area for future work.
\end{itemize}

\section{Conclusions}
\label{sec:conc}

\gls{ml} often requires sensitive data, but traditional encryption methods leave it vulnerable during processing. Homomorphic encryption, particularly the \gls{ckks} scheme, enables secure computations on encrypted data, preserving privacy throughout the \gls{ml} pipeline. This work presents the results of a proof-of-concept evaluating the feasibility of training models such as \gls{knn}, linear regression, and \gls{mlp} on encrypted datasets.

The experimental results demonstrate that homomorphic encryption enables accurate machine learning on encrypted data with minimal performance trade-offs. On the Boston Housing dataset, encrypted linear regression matched plaintext results (\gls{rmse}=4.67, R$^2$=0.71), while \gls{knn} (K=3) achieved \gls{rmse}=4.96, R$^2$=0.67 (both with <0.1\% decryption error). The \gls{mlp}, tested on encrypted data after plaintext training, reached R$^2$=0.69, though non-linear activations required simplification. Synthetic tests further confirmed accuracy (R$^2$>0.99). These findings prove that privacy-preserving \gls{ml} via Homomorphic encryption maintains model fidelity while securing sensitive data.

This work validates that machine learning models can be trained and deployed on encrypted data using polynomial approximations compatible with homomorphic encryption, ensuring end-to-end privacy without exposing raw data. By enabling secure computation on ciphertexts, the framework facilitates collaborative model training while mitigating privacy risks and legal barriers.

These techniques could encourage broader data sharing, fostering more accurate and generalizable models without compromising confidentiality. Although challenges such as efficiency and non-polynomial operations persist, the results establish a strong foundation for scalable, privacy-preserving machine learning. Future advancements may further expand its real-world applicability.

\begin{credits}
\subsubsection{\ackname} This work was supported and received funding from “CyberPRAISE - Cybersecurity research for PRivAte, Intelligent and truStablE solutions” - NORTE2030-FEDER-01820300.

\subsubsection{\discintname}
The authors have no competing interests to declare that are
relevant to the content of this article.
\end{credits}

%
%
%

\end{document}

%% file: glossary.tex
\newacronym{ml}{ML}{Machine Learning}
\newacronym{ckks}{CKKS}{Cheon-Kim-Kim-Song}
\newacronym{knn}{KNN}{K-Nearest Neighbors}
\newacronym{fhe}{FHE}{Fully Homomorphic Encryption}
\newacronym{he}{HE}{Homomorphic Encryption}
\newacronym{mlp}{MLP}{Multilayer Perceptron}
\newacronym{rmse}{RMSE}{Root Mean Square Error}
\newacronym{mae}{MAE}{Mean Absolute Error}
\newacronym{smc}{SMC}{Secure Multiparty Computation}